\documentclass[11pt]{article}
\usepackage{amssymb,amsmath,amsfonts}
\usepackage{graphicx}
\usepackage{graphics}
\usepackage{eepic,epsfig}

\begin{document}

\title{Fermionic Casimir effect for parallel plates in the presence of
compact dimensions with applications to nanotubes}
\author{ S. Bellucci$^{1}$\thanks{%
E-mail: bellucci@lnf.infn.it }\, and A. A. Saharian$^{2}$\thanks{%
E-mail: saharian@ictp.it } \\
\textit{$^1$ INFN, Laboratori Nazionali di Frascati,}\\
\textit{Via Enrico Fermi 40,00044 Frascati, Italy} \vspace{0.3cm}\\
\textit{$^2$ Department of Physics, Yerevan State University,}\\
\textit{1 Alex Manoogian Street, 0025 Yerevan, Armenia }}
\maketitle

\begin{abstract}
We evaluate the Casimir energy and force for a massive fermionic field in
the geometry of two parallel plates on background of Minkowski spacetime
with an arbitrary number of toroidally compactified spatial dimensions. The
bag boundary conditions are imposed on the plates and periodicity conditions
with arbitrary phases are considered along the compact dimensions. The
Casimir energy is decomposed into purely topological, single plate and
interaction parts. With independence of the lengths of the compact
dimensions and the phases in the periodicity conditions, the interaction
part of the Casimir energy is always negative. In order to obtain the
resulting force, the contributions from both sides of the plates must be
taken into account. Then, the forces coming from the topological parts of
the vacuum energy cancel out and only the interaction term contributes to
the Casimir force. Applications of the general formulae to Kaluza-Klein type
models and carbon nanotubes are given. In particular, we show that for
finite length metallic nanotubes the Casimir forces acting on the tube edges
are always attractive, whereas for semiconducting-type ones they are
attractive for small lengths of the nanotube and repulsive for large lengths.
\end{abstract}

\bigskip


\bigskip

\section{Introduction}

A key feature of most high energy theories of fundamental physics, including
supergravity and superstring theories, is the presence of compact spatial
dimensions. From an inflationary point of view universes with compact
spatial dimensions, under certain conditions, should be considered a rule
rather than an exception \cite{Lind04}. The models of a compact universe
with non-trivial topology may play an important role by providing proper
initial conditions for inflation (for physical motivations of considering
compact universes see also \cite{Star98}). There has been a large activity
to search for signatures of non-trivial topology by identifying ghost images
of galaxies, clusters or quasars. Recent progress in observations of the
cosmic microwave background provides an alternative way to observe the
topology of the universe \cite{Levi02}. If the scale of periodicity is close
to the particle horizon scale then the changed appearance of the microwave
background sky pattern offers a sensitive probe of the topology. An
interesting application of the field theoretical models with compact
dimensions recently appeared in nanophysics \cite{Sait98}. The
long-wavelength description of the electronic states in graphene can be
formulated in terms of the Dirac-like theory in 3-dimensional spacetime with
the Fermi velocity playing the role of speed of light (see, e.g., Refs. \cite%
{Seme84}). Single-walled carbon nanotubes are generated by rolling up a
graphene sheet to form a cylinder and the background spacetime for the
corresponding Dirac-like theory has topology $R^{2}\times S^{1}$.

In quantum field theory the boundary conditions imposed on fields along
compact dimensions change the spectrum of vacuum fluctuations. The resulting
energies and stresses are known as topological Casimir effect (for the
topological Casimir effect and its role in cosmology see \cite{Most97}-\cite%
{Duff86} and references therein). In the Kaluza-Klein-type models this
effect has been used as a stabilization mechanism for moduli fields which
parametrize the size and the shape of the extra dimensions. The Casimir
energy can also serve as a model of dark energy needed for the explanation
of the present accelerated expansion of the universe (see \cite{Milt03} and
references therein). In addition to its fundamental interest the Casimir
effect also plays an important role in the fabrication and operation of
nano- and micro-scale mechanical systems (see, for instance, \cite{CasNano}).

The effects of the toroidal compactification of spatial dimensions on the
properties of quantum vacuum for various spin fields have been discussed by
several authors (see, for instance, \cite{Most97}-\cite{Duff86}, \cite%
{CasTor,Saha08,Bell09} and references therein). The combined effect of extra
compactified dimensions and boundaries on the Casimir energy in the
classical configuration of two parallel plates has been recently considered
in \cite{Chen06} for a scalar field and in \cite{Popp04} for the
electromagnetic field. The Casimir energy and forces in braneworld models
have been evaluated in Refs.~\cite{Gold00} by using both dimensional and
zeta function regularization methods. Local Casimir densities in these
models were considered in Refs.~\cite{Knap04}. The Casimir effect in higher
dimensional generalizations of the Randall-Sundrum model with compact
internal spaces has been investigated in \cite{Flac03}. In the present
paper, we investigate the Casimir effect for a massive fermionic field in
the geometry of two parallel plates on background of spacetime with an
arbitrary number of toroidally compactified spatial dimensions. We will
assume generalized periodicity conditions along the compact dimensions with
arbitrary phases and MIT bag boundary conditions on the plates. This problem
in background of 4-dimensional Minkowski spacetime with trivial topology has
been considered in \cite{John75} for a massless field and in \cite{Mama80}
in the massive case (see also \cite{Most97}). For arbitrary number of
dimensions the corresponding results are generalized in Refs. \cite%
{Paol99,Eliz02} for the massless and massive cases respectively. The Casimir
problem for fermions coupled to a static background field in one spatial
dimension is investigated in \cite{Sund04}. The interaction energy density
and the force are computed in the limit that the background becomes
concentrated at two points. The fermionic Casimir effect for parallel plates
with imperfect bag boundary conditions modelled by $\delta $-like potentials
is studied in \cite{Fosc08}.

This paper is organized as follows. In the next section, we specify the
eigenfunctions and the eigenmodes for the Dirac equation in the region
between the plates assuming the bag boundary conditions on them. In section %
\ref{sec:CasEn}, by using the Abel-Plana-type summation formula, we present
the Casimir energy in the region between the plates as the sum of pure
topological, single plate and interaction parts. In section \ref%
{sec:CasForce} we consider the Casimir force acting on the plates. In
section \ref{sec:Zeta} we evaluate the Casimir energy and forces by making
use of the generalized zeta function technique. An alternative
representation of the single plate part of the Casimir energy is also given.
The special case of topology $R^{D-1}\times S^{1}$ is discussed in section %
\ref{sec:SpCase}. In section \ref{sec:Nanotubes} we give applications of
general formulae to the Casimir effect for electrons in finite-length carbon
nanotubes within the framework of 3-dimensional Dirac-like model. The main
results of the paper are summarized in section \ref{sec:Conclusion}.

\section{Eigenfunctions and eigenmodes}

\label{sec:Modes}

We consider a quantum fermionic field $\psi $ on background of $(D+1)$%
-dimensional flat spacetime with spatial topology $R^{p+1}\times (S^{1})^{q}$%
, $p+q+1=D$. The corresponding line element has the form
\begin{equation}
ds^{2}=dt^{2}-\sum_{l=1}^{D}(dz^{l})^{2},  \label{ds2}
\end{equation}%
where $-\infty <z^{l}<\infty $, $l=1,\ldots ,p+1$, and $0\leqslant
z^{l}\leqslant L_{l}$ for $l=p+2,\ldots ,D$. We assume that along the
compact dimensions the field obeys boundary conditions%
\begin{equation}
\psi (t,\mathbf{z}_{p},z^{p+1},\mathbf{z}_{q}+L_{l}\mathbf{e}_{l})=e^{2\pi
i\alpha _{l}}\psi (t,\mathbf{z}_{p},z^{p+1},\mathbf{z}_{q}),  \label{BC}
\end{equation}%
with constant phases $0\leqslant \alpha _{l}<1$. In (\ref{BC}), $\mathbf{z}%
_{p}=(z^{1},\ldots ,z^{p})$ and $\mathbf{z}_{q}=(z^{p+2},\ldots ,z^{D})$
denote the coordinates along uncompactified and compactified dimensions
respectively, $\mathbf{e}_{l}$\ is the unit vector along the direction of
the coordinate $z^{l}$, $l=p+2,\ldots ,D$. The periodicity conditions for
untwisted and twisted fermionic fields are obtained from (\ref{BC}) as
special cases with $\alpha _{l}=0$ and $\alpha _{l}=1/2$ respectively. As we
will see below, special cases $\alpha _{l}=0,1/3,2/3$ are realized in
nanotubes.

In this paper we are interested in the Casimir effect for the geometry of
two parallel plates placed at $z^{p+1}=0$ and $z^{p+1}=a$ on which the field
obeys the MIT bag boundary condition:
\begin{equation}
\left( 1+i\gamma ^{\mu }n_{\mu }\right) \psi =0\ ,\quad z^{p+1}=0,a,
\label{BagCond}
\end{equation}%
where $\gamma ^{\mu }$ are the Dirac matrices and $n_{\mu }$ is the normal
to the boundaries. In the $(D+1)$-dimensional spacetime the Dirac matrices
are $N\times N$ matrices with $N=2^{[(D+1)/2]}$, where the square brackets
mean the integer part of the enclosed expression. We will assume that these
matrices are given in the chiral representation:
\begin{equation}
\gamma ^{0}=\left(
\begin{array}{cc}
1 & 0 \\
0 & -1%
\end{array}%
\right) ,\;\gamma ^{\mu }=\left(
\begin{array}{cc}
0 & \sigma _{\mu } \\
-\sigma _{\mu }^{+} & 0%
\end{array}%
\right) ,\;\mu =1,2,\ldots ,D,  \label{gam0mu}
\end{equation}%
with the relation $\sigma _{\mu }\sigma _{\nu }^{+}+\sigma _{\nu }\sigma
_{\mu }^{+}=2\delta _{\mu \nu }$. In the discussion below we consider the
region between the plates, $0\leqslant z^{p+1}\leqslant a$, where we have $%
n_{\mu }=-\delta _{\mu }^{p+1}$ for the plate at $z^{p+1}=0$ and $n_{\mu
}=\delta _{\mu }^{p+1}$ for $z^{p+1}=a$.

The dynamics of the field is governed by the Dirac equation
\begin{equation}
i\gamma ^{\mu }\partial _{\mu }\psi -m\psi =0\ .  \label{Direq}
\end{equation}%
Assuming the time dependence in the form $e^{\pm i\omega t}$, the positive-
and negative-frequency solutions to this equation can be presented as%
\begin{eqnarray}
\psi _{\beta }^{(+)} &=&A_{\beta }e^{-i\omega t}\left(
\begin{array}{c}
\varphi \\
-i\boldsymbol{\sigma }^{+}\cdot \boldsymbol{\nabla }\varphi /\left( \omega
+m\right)%
\end{array}%
\right) ,  \notag \\
\psi _{\beta }^{(-)} &=&A_{\beta }e^{i\omega t}\left(
\begin{array}{c}
i\boldsymbol{\sigma }\cdot \boldsymbol{\nabla }\chi /\left( \omega +m\right)
\\
\chi%
\end{array}%
\right) ,  \label{Eigfunc}
\end{eqnarray}%
where $\boldsymbol{\sigma }=(\sigma _{1},\ldots ,\sigma _{D})$, $\omega =%
\sqrt{\mathbf{k}_{p}^{2}+k_{p+1}^{2}+\mathbf{k}_{q}^{2}+m^{2}}$, and%
\begin{eqnarray}
\varphi &=&e^{i\mathbf{k}_{\parallel }\cdot \mathbf{z}_{\parallel }}\left(
\varphi _{+}e^{ik_{p+1}z^{p+1}}+\varphi _{-}e^{-ik_{p+1}z^{p+1}}\right) ,
\notag \\
\chi &=&e^{-i\mathbf{k}_{\parallel }\cdot \mathbf{z}_{\parallel }}\left(
\chi _{+}e^{ik_{p+1}z^{p+1}}+\chi _{-}e^{-ik_{p+1}z^{p+1}}\right) ,
\label{phixi}
\end{eqnarray}%
with $\mathbf{k}_{\parallel }=(\mathbf{k}_{p},\mathbf{k}_{q})$ and $\mathbf{k%
}_{p}=(k_{1},\ldots ,k_{p})$, $\mathbf{k}_{q}=(k_{p+2},\ldots ,k_{D})$. The
eigenvalues for the components of the wave vector along the compactified
dimensions are determined from the periodicity conditions (\ref{BC}):%
\begin{equation}
\mathbf{k}_{q}=(2\pi (n_{p+2}+\alpha _{p+2})/L_{p+2},\ldots ,2\pi
(n_{D}+\alpha _{D})/L_{D}),\;n_{p+2},\ldots ,n_{D}=0,\pm 1,\pm 2,\ldots .
\label{kDn}
\end{equation}%
For the components along the uncompactified dimensions one has $-\infty
<k_{l}<\infty $, $l=1,\ldots ,p$.

From the boundary condition on the plate at $z^{p+1}=0$ we find the
following relations between the spinors in (\ref{phixi})%
\begin{eqnarray}
\varphi _{+} &=&-\frac{m(\omega +m)+k_{p+1}^{2}-k_{p+1}\sigma _{p+1}%
\boldsymbol{\sigma }_{\parallel }^{+}\cdot \mathbf{k}_{\parallel }}{%
(m-ik_{p+1})\left( \omega +m\right) }\varphi _{-},  \notag \\
\chi _{-} &=&-\frac{m(\omega +m)+k_{p+1}^{2}-k_{p+1}\sigma _{p+1}^{+}%
\boldsymbol{\sigma }_{\parallel }\cdot \mathbf{k}_{\parallel }}{%
(m+ik_{p+1})(\omega +m)}\chi _{+},  \label{phixiRel}
\end{eqnarray}%
where $\boldsymbol{\sigma }_{\parallel }=(\sigma _{1},\ldots ,\sigma
_{p},\sigma _{p+2},\ldots ,\sigma _{D})$. We will assume that they are
normalized in accordance with
\begin{equation}
\varphi _{-}^{+}\varphi _{-}=\chi _{+}^{+}\chi _{+}=1.  \label{normSpinor}
\end{equation}%
As a set of independent spinors we will take $\varphi _{-}=w^{(\sigma )}$
and $\chi _{+}=w^{(\sigma )\prime }$, where $w^{(\sigma )}$, $\sigma
=1,\ldots ,N/2$, are one-column matrices having $N/2$ rows with the elements
$w_{l}^{(\sigma )}=\delta _{l\sigma }$, and $w^{(\sigma )\prime
}=iw^{(\sigma )}$. Now the set of quantum numbers specifying the
eigenfunctions (\ref{Eigfunc}) is $\beta =(\mathbf{k},\sigma )$. From the
boundary condition at $z^{p+1}=a$ it follows that the eigenvalues of $%
k_{p+1} $ are roots of the transcendental equation
\begin{equation}
ma\sin (k_{p+1}a)/(k_{p+1}a)+\cos (k_{p+1}a)=0.  \label{kpvalues}
\end{equation}%
All these roots are real. We will denote the positive solutions of Eq. (\ref%
{kpvalues}) by $\lambda _{n}=k_{p+1}a$, $n=1,2,\ldots $. For a massless
field we have $\lambda _{n}=\pi (n-1/2)$. Note that the equation (\ref%
{kpvalues}) determining the eigenvalues for $k_{p+1}$ does not contain the
parameters of the compact subspace and is the same as in the corresponding
problem on the topologically trivial Minkowski spacetime (see \cite{Most97}).

The normalization coefficient $A_{\beta }$ in (\ref{Eigfunc}) is determined
from the orthonormalization condition%
\begin{equation}
\int d\mathbf{z}_{\parallel }\int_{0}^{a}dz^{p+1}\,\psi _{\beta }^{(\pm
)+}\psi _{\beta ^{\prime }}^{(\pm )}=\delta _{\beta \beta ^{\prime }},
\label{normaliz}
\end{equation}%
where $\delta _{\beta \beta ^{\prime }}$ is understood as the Dirac delta
function for continuous indices and the Kronecker delta for discrete ones.
Substituting the eigenfunctions (\ref{Eigfunc}) into this condition one finds%
\begin{equation}
A_{\beta }^{2}=\frac{\omega +m}{4(2\pi )^{p}\omega aV_{q}}\left[ 1-\frac{%
\sin (2k_{p+1}a)}{2k_{p+1}a}\right] ^{-1},  \label{normcoef}
\end{equation}%
where $V_{q}=L_{p+2}\cdots L_{D}$ is the volume of the compact subspace.

\section{Casimir energy}

\label{sec:CasEn}

For the spatial topology $R^{p+1}\times (S^{1})^{q}$ the vacuum energy (per
unit volume along the directions $z^{1},\ldots ,z^{p}$) in the region
between the plates is given by the following mode-sum:
\begin{equation}
E_{p+1,q}=-\frac{N}{2}\int \frac{d\mathbf{k}_{p}}{(2\pi )^{p}}\sum_{\mathbf{n%
}_{q}\in \mathbf{Z}^{q}}\sum_{n=1}^{\infty }\omega ,  \label{CasE}
\end{equation}%
where $\mathbf{n}_{q}=(n_{p+1},\ldots ,n_{D})$ and
\begin{equation}
\omega ^{2}=\mathbf{k}_{p}^{2}+k_{\mathbf{n}_{q}}^{2}+\lambda
_{n}^{2}/a^{2}+m^{2},\;k_{\mathbf{n}_{q}}^{2}=\sum_{l=p+2}^{D}[2\pi
(n_{l}+\alpha _{l})/L_{l}]^{2}.  \label{omega}
\end{equation}%
Of course, the expression on the right hand-side of Eq. (\ref{CasE}) is
divergent. We will assume that some cutoff function is present, without
writing it explicitly. For the further evaluation of the Casimir energy we
apply to the sum over $n$ in Eq. (\ref{CasE}) the Abel-Plana-like summation
formula
\begin{equation}
\sum_{n=1}^{\infty }\frac{\pi f(\lambda _{n})}{1-\sin (2\lambda
_{n})/(2\lambda _{n})}=-\frac{\pi maf(0)}{2(ma+1)}+\int_{0}^{\infty
}dzf(z)-i\int_{0}^{\infty }dt\frac{f(it)-f(-it)}{\frac{t+ma}{t-ma}e^{2t}+1}.
\label{SumForm}
\end{equation}%
This formula is obtained as a special case of the summation formula derived
in Ref. \cite{Rome02} by using the generalized Abel-Plana formula (see also
Ref. \cite{Saha08Rev}). Note that we have the relation%
\begin{equation}
1-\frac{\sin (2\lambda _{n})}{2\lambda _{n}}=1+\frac{ma}{(ma)^{2}+\lambda
_{n}^{2}}.  \label{rel1}
\end{equation}

By taking into account Eq. (\ref{rel1}), we apply the summation formula (\ref%
{SumForm}) with the function%
\begin{equation}
f(z)=\sqrt{z^{2}+\mathbf{k}_{p}^{2}a^{2}+m_{\mathbf{n}_{q}}^{2}a^{2}}\left[
1+\frac{ma}{(ma)^{2}+z^{2}}\right] ,  \label{fzforSum}
\end{equation}%
where we have introduced the notation%
\begin{equation}
m_{\mathbf{n}_{q}}^{2}=k_{\mathbf{n}_{q}}^{2}+m^{2}.  \label{kparal}
\end{equation}%
This allows to present the Casimir energy in the decomposed form%
\begin{equation}
E_{p+1,q}=aE_{p+1,q}^{(0)}+2E_{p+1,q}^{(1)}+\Delta E_{p+1,q},
\label{Edecomp}
\end{equation}%
where
\begin{equation}
E_{p+1,q}^{(0)}=-\frac{N}{2}\int \frac{d\mathbf{k}_{p+1}}{(2\pi )^{p+1}}%
\sum_{\mathbf{n}_{q}\in \mathbf{Z}^{q}}\sqrt{\mathbf{k}_{p+1}^{2}+m_{\mathbf{%
n}_{q}}^{2}},  \label{ERpS1}
\end{equation}%
is the Casimir energy (per unit volume along the directions $z^{1},\ldots
,z^{p+1}$) in the topology $R^{p+1}\times (S^{1})^{q}$ when the boundaries
are absent. The part
\begin{equation}
E_{p+1,q}^{(1)}=-\frac{N}{4\pi }\int \frac{d\mathbf{k}_{p}}{(2\pi )^{p}}%
\sum_{\mathbf{n}_{q}\in \mathbf{Z}^{q}}\left( -\frac{\pi }{2}\sqrt{%
k_{\parallel }^{2}+m^{2}}+m\int_{0}^{\infty }dz\frac{\sqrt{%
z^{2}+k_{\parallel }^{2}+m^{2}}}{m^{2}+z^{2}}\right) ,  \label{E1RpS1}
\end{equation}%
is the Casimir energy for a single plate (when the other plate is absent) in
the half-space. The last term in Eq. (\ref{Edecomp}),%
\begin{equation}
\Delta E_{p+1,q}=-\frac{N}{\pi }\int \frac{d\mathbf{k}_{p}}{(2\pi )^{p}}%
\sum_{\mathbf{n}_{q}\in \mathbf{Z}^{q}}\int_{\sqrt{\mathbf{k}_{p}^{2}+m_{%
\mathbf{n}_{q}}^{2}}}^{\infty }dz\frac{\sqrt{z^{2}-\mathbf{k}_{p}^{2}-m_{%
\mathbf{n}_{q}}^{2}}}{(z+m)e^{2az}+z-m}\left[ a(z-m)-\frac{m}{z+m}\right] ,
\label{DeltaE}
\end{equation}%
is the interaction part. This term is finite for all non-zero distances
between the plates and the cutoff function can be removed safely. Note that
the single plate part of the Casimir energy does not depend on the
separation between the plates and, hence, will not contribute to the Casimir
force.

The pure topological part (\ref{ERpS1}) is investigated in our previous
paper \cite{Bell09}. After the renormalization this part is presented in the
form%
\begin{equation}
E_{p+1,q}^{(0)}=2NV_{q}\sum_{j=p+2}^{D}\frac{(2\pi )^{-(j+1)/2}}{%
V_{D-j+1}L_{j}^{j}}\sum_{n=1}^{\infty }\frac{\cos (2\pi n\alpha _{j})}{%
n^{j+1}}\sum_{\mathbf{n}_{D-j}\in \mathbf{Z}^{D-j}}f_{(j+1)/2}(nL_{j}m_{%
\mathbf{n}_{D-j}}),  \label{Tmunutot}
\end{equation}%
where we have defined%
\begin{equation}
m_{\mathbf{n}_{D-j}}^{2}=\sum_{l=j+1}^{D}[2\pi (n_{l}+\alpha
_{l})/L_{l}]^{2}+m^{2}.  \label{omeganq}
\end{equation}%
Here and in the discussion below we use the notation
\begin{equation}
f_{\nu }(x)=x^{\nu }K_{\nu }(x).  \label{fnu}
\end{equation}%
An alternative expression for the topological part is obtained by using the
zeta function technique (see below). In particular, the topological part of
the Casimir energy is positive for untwisted fields ($\alpha _{l}=0$) and is
negative for twisted fields ($\alpha _{l}=1/2$).

\subsection{Single plate part}

Now let us consider the single plate part in the Casimir energy, given by
formula (\ref{E1RpS1}). First of all we note that this part vanishes for a
massless field. This is directly seen by taking into account that in the
limit $m\rightarrow 0$ the second term in braces of (\ref{E1RpS1}) gives
nonzero contribution which cancels the first term. Another way to see this
is the following. For a massless field $\lambda _{n}=\pi (n-1/2)$ and we can
apply to the corresponding sum in the Casimir energy (\ref{CasE}) the
Abel-Plana formula in the form (see, \cite{Most97,Saha08Rev})%
\begin{equation}
\sum_{n=1}^{\infty }f(n-1/2)=\int_{0}^{\infty }dx\,f(x)-i\int_{0}^{\infty }dx%
\frac{f(ix)-f(-ix)}{e^{2\pi x}+1}.  \label{AbelPlana}
\end{equation}%
The part of the vacuum energy with the first term on the right of this
formula gives the topological Casimir energy $E_{p+1,q}^{(0)}$ and the
second term corresponds to the interaction part $\Delta E_{p+1,q}$.

For the further evaluation of the single plate part for a massive field we
apply to the sum over $n_{p+2}$ in Eq. (\ref{E1RpS1}) the Abel-Plana
summation formula in the form \cite{Inui03}%
\begin{equation}
\sum_{n_{p+2}=-\infty }^{+\infty }f(|n_{p+2}+\alpha
_{p+2}|)=2\int_{0}^{\infty }dx\,f(x)+i\int_{0}^{\infty }dx\,\sum_{\lambda
=\pm 1}\frac{f(ix)-f(-ix)}{e^{2\pi (x+i\lambda \alpha _{p+2})}-1}.
\label{sumform1}
\end{equation}%
The part with the first term on the right of this formula gives the Casimir
energy for a single plate in the case of topology $R^{p+2}\times
(S^{1})^{q-1}$ and we obtain the following recurrence formula%
\begin{equation}
\varepsilon _{p+1,q}^{(1)}=\varepsilon _{p+2,q-1}^{(1)}+\Delta
_{p+2}\varepsilon _{p+1,q}^{(1)},  \label{E1RpS1Rec}
\end{equation}%
where $m_{\mathbf{n}_{q-1}}=\sqrt{k_{\mathbf{n}_{q-1}}^{2}+m^{2}}$ and we
have introduced the vacuum energy per unit volume of the compact subspace $%
\varepsilon _{p+1,q}^{(1)}=E_{p+1,q}^{(1)}/V_{q}$. In (\ref{E1RpS1Rec}),
\begin{eqnarray}
\Delta _{p+2}\varepsilon _{p+1,q}^{(1)} &=&-\frac{2NL_{p+2}}{(2\pi
)^{p/2+2}V_{q}}\sum_{n=1}^{\infty }\frac{\cos (2\pi n\alpha _{p+2})}{%
(nL_{p+2})^{p+2}}\sum_{\mathbf{n}_{q-1}\in \mathbf{Z}^{q-1}}\left[ \frac{\pi
}{2}f_{p/2+1}(nL_{p+2}m_{\mathbf{n}_{q-1}})\right.  \notag \\
&&\left. -\int_{m_{\mathbf{n}_{q-1}}}^{\infty }dx\frac{m}{x^{2}-k_{\mathbf{n}%
_{q-1}}^{2}}\frac{xf_{p/2+1}(nL_{p+2}x)}{\sqrt{x^{2}-m_{\mathbf{n}_{q-1}}^{2}%
}}\right] ,  \label{DeltE1RpS1}
\end{eqnarray}%
is the part induced by the compactness of the direction $z^{p+2}$. In
deriving this formula we have used the integration formulae%
\begin{equation}
\int d\mathbf{k}_{p}\int_{\sqrt{\mathbf{k}_{p}^{2}+c^{2}}}^{\infty }dz(z^{2}-%
\mathbf{k}_{p}^{2}-c^{2})^{(s+1)/2}f(z)=\frac{\pi ^{p/2}\Gamma ((s+3)/2)}{%
\Gamma ((p+s+3)/2)}\int_{c}^{\infty }dx\,x(x^{2}-c^{2})^{(p+s+1)/2}f(x),
\label{intform3}
\end{equation}%
and%
\begin{equation}
\sum_{\lambda =\pm 1}\int_{b}^{\infty }dx\frac{(x^{2}-b^{2})^{(p+1)/2}}{%
e^{L_{p+2}x+2\pi i\lambda \alpha _{p+2}}-1}=\sum_{n=1}^{\infty }\frac{%
2^{p/2+2}\Gamma ((p+3)/2)}{\sqrt{\pi }(nL_{p+2})^{p+2}}\cos (2\pi n\alpha
_{p+2})f_{p/2+1}(nL_{p+2}b).  \label{intform2}
\end{equation}%
The first of these formulae is obtained by integrating over the angular part
of $\mathbf{k}_{p}$, changing the integration variable to $y=\sqrt{z^{2}-%
\mathbf{k}_{p}^{2}-c^{2}}$, and introducing polar coordinates in the $(|%
\mathbf{k}_{p}|,y)$-plane. Formula (\ref{intform2}) is obtained expanding
the integrand by using the relation $(e^{u}-1)^{-1}=\sum_{n=1}^{\infty
}e^{-nu}$ and integrating the separate terms in this expansion.

After the recurring application of formula (\ref{E1RpS1Rec}) the Caimir
energy for a single plate is presented in the form%
\begin{equation}
E_{p+1,q}^{(1)}=V_{q}E_{D,0}^{(1)}+E_{p+1,q}^{(1,c)},  \label{epsdecomp}
\end{equation}%
where $E_{D,0}^{(1)}$ is the Casimir energy per unit volume along the
directions $z^{1},\ldots ,z^{D-1}$ for a single plate in Minkowski spacetime
with trivial topology and the second term,%
\begin{equation}
E_{p+1,q}^{(1,c)}=V_{q}\sum_{j=p+2}^{D}\Delta _{j}\varepsilon
_{j-1,D+1-j}^{(1)},  \label{E1ca}
\end{equation}
is the topological part. The latter is finite and in the corresponding
expression the cutoff function can be removed. The renormalization is needed
for the term $E_{D,0}^{(1)}$ only.

\subsection{Interaction part}

By using Eq. (\ref{intform3}), the interaction part of the Casimir energy is
presented in the form%
\begin{equation}
\Delta E_{p+1,q}=-\frac{(4\pi )^{-(p+1)/2}N}{\Gamma ((p+3)/2)}\sum_{\mathbf{n%
}_{q}\in \mathbf{Z}^{q}}\int_{m_{\mathbf{n}_{q}}}^{\infty }dz\frac{(z^{2}-m_{%
\mathbf{n}_{q}}^{2})^{(p+1)/2}}{(z+m)e^{2az}+z-m}\left[ a(z-m)-\frac{m}{z+m}%
\right] .  \label{DeltaE1}
\end{equation}%
From here it follows that this part is always negative and it is a
monotonically increasing function of $a$. By taking into account the relation%
\begin{equation}
\frac{a(z-m)-m/(z+m)}{(z+m)e^{2az}+z-m}=-\frac{1}{2}\frac{d}{dz}\ln \left( 1+%
\frac{z-m}{z+m}e^{-2az}\right) ,  \label{rel2}
\end{equation}%
and integrating by parts, Eq. (\ref{DeltaE1}) is written in the equivalent
form%
\begin{equation}
\Delta E_{p+1,q}=-\frac{(4\pi )^{-(p+1)/2}N}{\Gamma ((p+1)/2)}\sum_{\mathbf{n%
}_{q}\in \mathbf{Z}^{q}}\int_{m_{\mathbf{n}_{q}}}^{\infty }dz\,z(z^{2}-m_{%
\mathbf{n}_{q}}^{2})^{(p-1)/2}\ln \left( 1+\frac{z-m}{z+m}e^{-2az}\right) .
\label{DeltaE2}
\end{equation}%
For a massless fermionic field from here we find%
\begin{eqnarray}
\Delta E_{p+1,q} &=&-a\frac{(4\pi )^{-(p+1)/2}N}{\Gamma ((p+3)/2)}\sum_{%
\mathbf{n}_{q}\in \mathbf{Z}^{q}}\int_{k_{\mathbf{n}_{q}}}^{\infty }dz\frac{%
(z^{2}-k_{\mathbf{n}_{q}}^{2})^{(p+1)/2}}{e^{2az}+1}  \notag \\
&=&\frac{(2\pi )^{-p/2-1}N}{(2a)^{p+1}}\sum_{\mathbf{n}_{q}\in \mathbf{Z}%
^{q}}\sum_{n=1}^{\infty }\frac{(-1)^{n}}{n^{p+2}}f_{p/2+1}(2ank_{\mathbf{n}%
_{q}}),  \label{DeltaEm0}
\end{eqnarray}%
where the function $f_{\nu }(x)$ is defined by Eq. (\ref{fnu}).

Let us consider the asymptotic behavior of the interaction part in the
Casimir energy at small and large separations between the plates. In the
limit $L_{l}\gg a$ the main contribution comes from large values of $n_{l}$,
$l=p+2,\cdots ,D$, and we can replace the summation by the integration: $%
\sum_{\mathbf{n}_{q}\in \mathbf{Z}^{q}}\rightarrow \int d\mathbf{n}_{q}$. By
making use of the integration formula (\ref{intform3}) with $p\rightarrow q$%
, we find%
\begin{equation}
\Delta E_{p+1,q}\approx V_{q}\Delta E_{D,0}=-V_{q}\frac{(4\pi )^{-D/2}N}{%
\Gamma (D/2)}\int_{m}^{\infty }dzz(z^{2}-m^{2})^{D/2-1}\ln \left( 1+\frac{z-m%
}{z+m}e^{-2az}\right) ,  \label{DeltaEsmalla}
\end{equation}%
where $\Delta E_{D,0}$ is the interaction part of the fermionic Casimir
energy per unit volume along the directions $z^{1},\ldots ,z^{D-1}$ for two
parallel plates in $D$-dimensional space with trivial topology (see Refs.
\cite{Most97,Mama80} for the case $D=3$ and Ref. \cite{Eliz02} for general $%
D $). Note that for a massless field we have%
\begin{equation}
\Delta E_{D,0}=-\frac{N(1-2^{-D})}{(4\pi )^{(D+1)/2}a^{D}}\Gamma
((D+1)/2)\zeta (D+1),  \label{DED0}
\end{equation}%
where $\zeta (x)$ is the Riemann zeta function.

Now let us consider the limit $L_{l}\ll a$. In this case and for $\alpha
_{l}=0$ the main contribution comes from the zero mode with $\mathbf{n}%
_{q}=0 $ and $\Delta E_{p+1,q}/N$ coincides with the corresponding result
for the Casimir effect in topologically trivial $(p+1)$-dimensional space:%
\begin{eqnarray}
\Delta E_{p+1,q} &\approx &\frac{N}{N_{p}}\Delta E_{p+1,0}=-\frac{(4\pi
)^{-(p+1)/2}N}{\Gamma ((p+1)/2)}\int_{m}^{\infty }dz\,z  \notag \\
&&\times (z^{2}-m^{2})^{(p-1)/2}\ln \left( 1+\frac{z-m}{z+m}e^{-2az}\right) ,
\label{DeltaElargea}
\end{eqnarray}%
where $N_{p}=2^{[(p+1)/2]}$.\ The contribution of the nonzero modes is
exponentially suppressed. For $\alpha _{l}\neq 0$ the zero mode is absent
and assuming that $am$ is fixed, to the leading order we have%
\begin{equation}
\Delta E_{p+1,q}\approx -\frac{Ne^{-2ac_{0}}}{2(4\pi a)^{(p+1)/2}}%
c_{0}^{(p+1)/2},  \label{DeltaElargea1}
\end{equation}%
where
\begin{equation}
c_{0}^{2}=\sum_{l=p+2}^{D}(2\pi \beta _{l}a/L_{l})^{2},\;\beta _{l}=\min
(\alpha _{l},1-\alpha _{l}).  \label{c0}
\end{equation}%
In this case the interaction part of the Casimir energy is exponentially
suppressed.

In the discussion above we have considered the region between the plates.
The plates divide the background space into three regions: $z^{p+1}<0$, $%
0<z^{p+1}<a$, and $z^{p+1}>a$. The vacuum energy in the regions $z^{p+1}<0$
and $z^{p+1}>a$ is obtained from the results given above in the limit $%
a\rightarrow \infty $. In this limit the interaction part vanishes and we
have%
\begin{equation}
E_{p+1,q}=aE_{p+1,q}^{(0)}+E_{p+1,q}^{(1)},\;z^{p+1}<0,\;z^{p+1}>a,
\label{EcasSing}
\end{equation}%
with the topological and single plate parts given by Eqs. (\ref{Tmunutot})
and (\ref{epsdecomp}).

\section{The Casimir force}

\label{sec:CasForce}

The total vacuum energy in the region $0\leqslant z^{l}\leqslant c_{l}$, $%
l=1,\ldots ,p$, $0\leqslant z^{p+1}\leqslant a$ will be $E_{p+1,q}c_{1}%
\cdots c_{p}$ and the volume of this region is $V=c_{1}\cdots c_{p}aV_{q}$.
The vacuum stress at $z^{p+1}=0+$ is given by%
\begin{equation}
P_{p+1,q}(0+)=-\frac{\partial }{\partial V}E_{p+1,q}c_{1}\cdots
c_{p}=P_{p+1,q}^{(0)}+\Delta P_{p+1,q},  \label{P}
\end{equation}%
where we have introduced the notations%
\begin{equation}
P_{p+1,q}^{(0)}=-\frac{E_{p+1,q}^{(0)}}{V_{q}},\;\Delta P_{p+1,q}=-\frac{1}{%
V_{q}}\frac{\partial }{\partial a}\Delta E_{p+1,q}.  \label{P0DeltP}
\end{equation}%
The vacuum stress at $z^{p+1}=a-$ is given by the same expression. The term $%
P_{p+1,q}^{(0)}$ does not depend on the separation between the plates and is
the pure topological part of the vacuum force. The term $\Delta P_{p+1,q}$
is induced by the presence of the second plate and determines the
interaction force between the plates. Using the formula for $\Delta
E_{p+1,q} $, for this part we find%
\begin{equation}
\;\Delta P_{p+1,q}=-\frac{2(4\pi )^{-(p+1)/2}N}{\Gamma ((p+1)/2)V_{q}}\sum_{%
\mathbf{n}_{q}\in \mathbf{Z}^{q}}\int_{m_{\mathbf{n}_{q}}}^{\infty }dz\frac{%
z^{2}(z^{2}-m_{\mathbf{n}_{q}}^{2})^{(p-1)/2}}{\frac{z+m}{z-m}e^{2az}+1}.
\label{DeltP}
\end{equation}%
Now we see that $\Delta P_{p+1,q}<0$ independent of the boundary conditions
imposed on the field along the compactified dimensions and, hence, the
interaction forces between the plates are always attractive. For a massless
fermionic field we have%
\begin{equation}
\;\Delta P_{p+1,q}=-\frac{2N}{(2\pi )^{p/2+1}V_{q}}\sum_{\mathbf{n}_{q}\in
\mathbf{Z}^{q}}\sum_{n=1}^{\infty }(-1)^{n}\frac{f_{p/2+1}(2ank_{\mathbf{n}%
_{q}})-f_{p/2+2}(2ank_{\mathbf{n}_{q}})}{(2an)^{p+2}}.  \label{DeltaPm0}
\end{equation}

For small separations between the plates, $L_{l}\gg a$, we replace the
summation over $\mathbf{n}_{q}$\ by the integration. In the way similar to
that we have used for the Casimir energy, it can be seen that in the leading
order the interaction force coincides with the corresponding result for
parallel plates on background of $D$-dimensional space with trivial topology:%
\begin{equation}
\;\Delta P_{p+1,q}\approx \Delta P_{D,0}=-\frac{2N}{(4\pi )^{D/2}\Gamma (D/2)%
}\int_{m}^{\infty }dz\frac{z^{2}(z^{2}-m^{2})^{D/2-1}}{\frac{z+m}{z-m}%
e^{2az}+1}.  \label{DPsmall}
\end{equation}%
The contribution of the nonzero modes is exponentially small. For the
massless field we have%
\begin{equation}
\;\Delta P_{D,0}=-\frac{ND(1-2^{-D})}{(4\pi )^{(D+1)/2}a^{D+1}}\Gamma
((D+1)/2)\zeta (D+1).  \label{DPm0}
\end{equation}%
This result can also be directly obtained from Eq. (\ref{DED0}).

For large inter-plate separations, $L_{l}\ll a$, and for $\alpha _{l}=0$ the
main contribution comes from the zero mode $\mathbf{n}_{q}=0$ and $%
V_{q}\Delta P_{p+1,q}/N$ coincides with the corresponding result for the
Casimir effect in $(p+1)$-dimensional space:%
\begin{equation}
\;\Delta P_{p+1,q}\approx \frac{N}{N_{p}V_{q}}\Delta P_{p+1,0}=-\frac{2(4\pi
)^{-(p+1)/2}N}{\Gamma ((p+1)/2)V_{q}}\int_{m}^{\infty }dz\frac{%
z^{2}(z^{2}-m^{2})^{(p-1)/2}}{\frac{z+m}{z-m}e^{2az}+1}.  \label{DPlarge}
\end{equation}%
If $\alpha _{l}\neq 0$ and $am$ is fixed the interaction force is
exponentially suppressed:%
\begin{equation}
\;\Delta P_{p+1,q}\approx -\frac{Nc_{0}^{(p+3)/2}e^{-2ac_{0}}}{(4\pi
a)^{(p+1)/2}V_{q}},  \label{DPlarge1}
\end{equation}%
with $c_{0}$ defined by Eq. (\ref{c0}).

If the quantum field lives in all regions, in considering the total forces
acting on the plate we should also take into account the force acting on the
sides $z^{p+1}=0-$ and $z^{p+1}=a+$. The corresponding forces per unit
surface are equal to $P_{p+1,q}^{(0)}$ and they are directed along the
positive/negative direction of the axis $z^{p+1}$ in the case $%
P_{p+1,q}^{(0)}>0$/$P_{p+1,q}^{(0)}<0$. Now we see that the topological
parts of the force acting from the left and right sides of the plate
compensate and the resulting force is determined by (\ref{DeltP}) and it is
attractive for all inter-plate separations. There are physical situations
[bag model, finite length carbon nanotubes (see below)], where the quantum
field is confined to the interior of some region and there is no field
outside. For the problem under consideration, if the quantum field is
confined in the region between the plates, the total Casimir force acting
per unit surface of the plate is determined by Eq. (\ref{P}) and the pure
topological part contributes as well. At large distances this part dominates
and the corresponding forces tend to increase/decrease the distance between
the plates when $P_{p+1,q}^{(0)}>0$/$P_{p+1,q}^{(0)}<0$. In particular, $%
P_{p+1,q}^{(0)}<0$ for untwisted fields and $P_{p+1,q}^{(0)}>0$ for twisted
fields. Hence, if the quantum field is confined in the region between the
plates, for untwisted fields the Casimir forces are attractive for all
separations. For twisted fields these forces are attractive for small
distances and they are repulsive at large distances.

\section{Zeta function approach}

\label{sec:Zeta}

In this section, for the evaluation of the vacuum energy in the region $%
0\leqslant z^{p+1}\leqslant a$ we will use the zeta function technique \cite%
{Eliz94,Kirs01}. This allows to obtain alternative representations for the
pure topological and single plate parts in the Casimir effect. Instead of
the divergent expression on the right of Eq. (\ref{CasE}) we consider the
finite quantity
\begin{equation}
E_{p+1,q}(\mu ,s)=-\mu ^{2s+1}\frac{N}{2}\int \frac{d\mathbf{k}_{p}}{(2\pi
)^{p}}\sum_{\mathbf{n}_{q}\in \mathbf{Z}^{q}}\sum_{n=1}^{\infty }\left(
\mathbf{k}_{p}^{2}+m_{\mathbf{n}_{q}}^{2}+\lambda _{n}^{2}/a^{2}\right)
^{-s},  \label{Es}
\end{equation}%
where the arbitrary mass scale $\mu $ is introduced in order to keep the
dimensionality of the expression. Performing the integration over $\mathbf{k}%
_{p}$, we find%
\begin{equation}
E_{p+1,q}(\mu ,s)=-\mu ^{2s+1}\frac{N\Gamma (s-p/2)}{2(4\pi )^{p/2}\Gamma (s)%
}\sum_{\mathbf{n}_{q}\in \mathbf{Z}^{q}}\sum_{n=1}^{\infty }(m_{\mathbf{n}%
_{q}}^{2}+\lambda _{n}^{2}/a^{2})^{p/2-s}.  \label{Es1}
\end{equation}

The computation of the Casimir energy requires the analytic continuation of $%
E_{p+1,q}(\mu ,s)$ to the value $s=-1/2$. The starting point of our
consideration is the representation of the partial zeta function as a
contour integral in the complex plane $z$:%
\begin{equation}
\sum_{n=1}^{\infty }(m_{\mathbf{n}_{q}}^{2}+\lambda _{n}^{2}/a^{2})^{p/2-s}=%
\frac{1}{2\pi i}\int_{C}dz\,(z^{2}/a^{2}+m_{\mathbf{n}_{q}}^{2})^{p/2-s}%
\frac{d}{dz}\ln \left[ \frac{(ma/z)\sin z+\cos z}{1+ma}\right] ,
\label{zetaint}
\end{equation}%
where $C$ denotes a closed counterclockwise contour enclosing all zeros $%
\lambda _{n}$. We assume that the contour $C$ is made of a large semicircle
(with radius tending to infinity) centered at the origin and placed to its
right, plus a straight part overlapping the imaginary axis and avoiding the
points $\pm im_{\mathbf{n}_{q}}a$ by small semicircles in the right
half-plane. When the radius of the large semicircle tends to infinity the
corresponding contribution vanishes for ${\mathrm{Re}}\,s>(p+1)/2$. Assuming
that $(p+1)/2<{\mathrm{Re}}\,s<p/2+1$, from (\ref{zetaint}) we find the
following integral representation for the partial zeta function:
\begin{eqnarray}
\sum_{n=1}^{\infty }(m_{\mathbf{n}_{q}}^{2}+\lambda _{n}^{2}/a^{2})^{p/2-s}
&=&\frac{1}{\pi }\sin [\pi (s-p/2)]\int_{m_{\mathbf{n}_{q}}}^{\infty
}dz\,(z^{2}-m_{\mathbf{n}_{q}}^{2})^{p/2-s}  \notag \\
&&\times \frac{d}{dz}\ln \left[ \frac{(m/z)\sinh (az)+\cosh (az)}{1+ma}%
\right] .  \label{zetaint1}
\end{eqnarray}

Hence, the regularized vacuum energy is presented in the form%
\begin{eqnarray}
E_{p+1,q}(\mu ,s) &=&-\frac{(4\pi )^{-p/2}\mu ^{2s+1}N}{2\Gamma (s)\Gamma
(1-s+p/2)}\sum_{\mathbf{n}_{q}\in \mathbf{Z}^{q}}\int_{m_{\mathbf{n}%
_{q}}}^{\infty }dz\,(z^{2}-m_{\mathbf{n}_{q}}^{2})^{p/2-s}  \notag \\
&&\times \frac{d}{dz}\ln \left[ \frac{(m/z)\sinh (az)+\cosh (az)}{1+ma}%
\right] .  \label{Es2}
\end{eqnarray}%
Now we decompose the logarithmic term in this expression as%
\begin{equation}
\frac{d}{dz}\ln \left[ \frac{(m/z)\sinh (az)+\cosh (az)}{1+ma}\right] =a+%
\frac{d}{dz}\ln (1+m/z)+\frac{d}{dz}\ln \left( 1+\frac{z-m}{z+m}%
e^{-2az}\right) .  \label{decln}
\end{equation}%
As a result, we have the following decomposition of the generalized zeta
function:
\begin{equation}
E_{p+1,q}(\mu ,s)=aE_{p+1,q}^{(0)}(\mu ,s)+2E_{p+1,q}^{(1)}(\mu ,s)+\Delta
E_{p+1,q}(\mu ,s),  \label{Esdecomp}
\end{equation}%
where the first, second and third terms on the right hand-side come from the
corresponding terms in Eq. (\ref{decln}). The interaction term $\Delta E(\mu
,s)$ in Eq. (\ref{Esdecomp}) is finite at the physical point $s=-1/2$ and
gives the result (\ref{DeltaE2}): $\Delta E_{p+1,q}(\mu ,-1/2)=\Delta
E_{p+1,q}$. Below we will be focused on the pure topological and single
plate parts.

First let us consider the term $E_{p+1,q}^{(0)}(\mu ,s)$. This term is the
regularized vacuum energy in the topology $R^{p+1}\times (S^{1})^{q}$
without boundaries. In this term the integration over $z$ is done explicitly
and we find%
\begin{equation}
E_{p+1,q}^{(0)}(\mu ,s)=-\frac{\mu ^{2s+1}N\Gamma (s-(p+1)/2)}{2(4\pi
)^{(p+1)/2}\Gamma (s)}\sum_{\mathbf{n}_{q}\in \mathbf{Z}^{q}}m_{\mathbf{n}%
_{q}}^{p+1-2s}.  \label{Etops}
\end{equation}%
Further analytic continuation of this expression to the physical point $%
s=-1/2$ is done by using the extended Chowla--Selberg formula \cite{Eliz98}
and the corresponding result is given by the expression \cite{Bell09}:%
\begin{equation}
E_{p+1,q}^{(0)}=\frac{Nm^{D+1}V_{q}}{(2\pi )^{(D+1)/2}}\sideset{}{'}{\sum}_{%
\mathbf{m}_{q}\in \mathbf{Z}^{q}}\cos (2\pi \mathbf{m}_{q}\cdot %
\boldsymbol{\alpha }_{q})\frac{f_{(D+1)/2}(mg(\mathbf{L}_{q},\mathbf{m}_{q}))%
}{(mg(\mathbf{L}_{q},\mathbf{m}_{q}))^{D+1}},  \label{T00zeta}
\end{equation}%
where we have used the notation%
\begin{equation}
g(\mathbf{L}_{q},\mathbf{m}_{q})=\left(
\sum_{i=p+2}^{D}L_{i}^{2}m_{i}^{2}\right) ^{1/2}.  \label{gLm}
\end{equation}%
The prime on the summation sign in Eq. (\ref{T00zeta}) means that the term $%
\mathbf{m}_{q}=0$ should be excluded from the sum.

Now we turn to the part $E_{p+1,q}^{(1)}(\mu ,s)$ which is the regularized
vacuum energy in the half-space induced by a single plate. In the
corresponding integral representation we expand $\ln (1+m/z)$ in powers of $%
m/z$ and integrate over $z$ explicitly. This leads to the result%
\begin{equation}
E_{p+1,q}^{(1)}(\mu ,s)=-\frac{\mu ^{2s+1}N}{8(4\pi )^{p/2}\Gamma (s)}%
\sum_{l=1}^{\infty }(-1)^{l}m^{l}\frac{\Gamma ((l-p)/2+s)}{\Gamma (l/2+1)}%
\sum_{\mathbf{n}_{q}\in \mathbf{Z}^{q}}m_{\mathbf{n}_{q}}^{p-2s-l}.
\label{E1s}
\end{equation}%
The application to the multiseries over $\mathbf{n}_{q}$ of the extended
Chowla--Selberg formula allows to present $E_{p+1,q}^{(1)}(\mu ,s)$ as the
sum of two parts:%
\begin{equation}
E_{p+1,q}^{(1)}(\mu ,s)=V_{q}E_{R^{D}}^{(1)}(\mu ,s)+E_{p+1,q}^{(1,c)}(\mu
,s),  \label{E1sdecomp}
\end{equation}%
where
\begin{equation}
E_{R^{D}}^{(1)}(\mu ,s)=-\frac{\mu ^{2s+1}Nm^{D-2s-1}}{8(4\pi
)^{(D-1)/2}\Gamma (s)}\sum_{l=1}^{\infty }(-1)^{l}\frac{\Gamma (s+(l+1-D)/2)%
}{\Gamma (l/2+1)},  \label{E1RD}
\end{equation}%
is the corresponding quantity in the case of trivial topology $R^{D}$. The
topological term $E_{p+1,q}^{(1,c)}(\mu ,s)$ is finite at the physical point
$s=-1/2$ and the topological part of the vacuum energy for a single plate
has the form%
\begin{eqnarray}
E_{p+1,q}^{(1,c)} &=&E_{p+1,q}^{(1,c)}(\mu ,-1/2)=\frac{Nm^{D}V_{q}}{4(2\pi
)^{D/2}}\sum_{l=1}^{\infty }\frac{2^{-l/2}(-1)^{l}}{\Gamma (l/2+1)}  \notag
\\
&&\times \sideset{}{'}{\sum}_{\mathbf{m}_{q}\in \mathbf{Z}^{q}}\cos (2\pi
\mathbf{m}_{q}\cdot \boldsymbol{\alpha }_{q})f_{(l-D)/2}(mg(\mathbf{L}_{q},%
\mathbf{m}_{q})),  \label{E1c}
\end{eqnarray}%
where we have used the relation $f_{-\nu }(x)=x^{-2\nu }f_{\nu }(x)$. Note
that we can write the function $\cos (2\pi \mathbf{m}_{q}\cdot %
\boldsymbol{\alpha }_{q})$ on the right of formula (\ref{E1c}) in the form
of the product $\prod_{i=p+2}^{D}\cos (2\pi m_{i}\alpha _{i})$. The
equivalence of two representations (\ref{epsdecomp}) and (\ref{E1c}) for the
topological part in the Casimir energy for a single plate can be seen by
making use of the relation \cite{Bell09}%
\begin{eqnarray}
&&\sum_{\mathbf{m}_{q-1}\in Z^{q-1}}\cos (2\pi \mathbf{m}_{q-1}\cdot %
\boldsymbol{\alpha }_{q-1})f_{(l-D)/2}(mg(\mathbf{L}_{q},\mathbf{m}_{q}))
\notag \\
&& =\frac{(2\pi )^{(q-1)/2}L_{p+2}}{V_{q}m^{D-l}}\sum_{\mathbf{n}_{q-1}\in
\mathbf{Z}^{q-1}}\frac{f_{(p-l)/2+1}(m_{p+2}L_{p+2}m_{\mathbf{n}_{q-1}})}{%
(m_{p+2}L_{p+2})^{p-l+2}},  \label{rel3}
\end{eqnarray}%
and the formula%
\begin{equation}
\frac{2}{\pi }\int_{1}^{\infty }dx\frac{c}{x^{2}-1+c^{2}}\frac{xf_{p/2+1}(bx)%
}{\sqrt{x^{2}-1}}=\sum_{l=0}^{\infty }\frac{2^{-l/2}(-1)^{l}}{\Gamma (l/2+1)}%
(bc)^{l}f_{(p-l)/2+1}(b),  \label{rel4}
\end{equation}%
valid for $0\leqslant c\leqslant 1$.

\section{Special case of topology}

\label{sec:SpCase}

By taking into account the importance of special case $p=D-2$, $q=1$ in
Kaluza-Klein models and in carbon nanotubes, in this section we consider it
separately. For the later convenience, the parameters of the compactified
dimension we will denote by $L_{D}=L$ and $\alpha _{D}=\alpha $. The
corresponding formulae for the separate parts in the Casimir energy take the
form%
\begin{eqnarray}
E_{D-1,1}^{(0)} &=&\frac{2NL^{-D}}{(2\pi )^{(D+1)/2}}\sum_{n=1}^{\infty }%
\frac{\cos (2\pi n\alpha )}{n^{D+1}}f_{(D+1)/2}(mnL),  \notag \\
E_{D-1,1}^{(1,c)} &=&\frac{Nm^{D}L}{2(2\pi )^{D/2}}\sum_{n=1}^{\infty }\cos
(2\pi n\alpha )\sum_{l=1}^{\infty }\frac{2^{-l/2}(-1)^{l}}{\Gamma (l/2+1)}%
f_{(l-D)/2}(mnL),  \label{Especial} \\
\Delta E_{D-1,1} &=&-\frac{(4\pi )^{-(D-1)/2}N}{\Gamma ((D-1)/2)}%
\sum_{l=-\infty }^{+\infty }\int_{m_{l}}^{\infty
}dzz(z^{2}-m_{l}^{2})^{(D-3)/2}\ln \left( 1+\frac{z-m}{z+m}e^{-2az}\right) ,
\notag
\end{eqnarray}%
where we have introduced the notation%
\begin{equation}
m_{l}^{2}=[2\pi (l+\alpha )/L]^{2}+m^{2}.  \label{ml}
\end{equation}%
An equivalent representation of the single plate part is obtained from Eq. (%
\ref{E1ca}):%
\begin{equation}
E_{D-1,1}^{(1,c)}=-\frac{2NL}{(2\pi )^{D/2+1}}\sum_{n=1}^{\infty }\frac{\cos
(2\pi n\alpha )}{(nL)^{D}}\left[ \frac{\pi }{2}f_{D/2}(nLm)-\int_{1}^{\infty
}dx\frac{f_{D/2}(nLmx)}{x\sqrt{x^{2}-1}}\right] .  \label{E1plSpecial}
\end{equation}

\begin{figure}[tbph]
\begin{center}
\begin{tabular}{cc}
\epsfig{figure=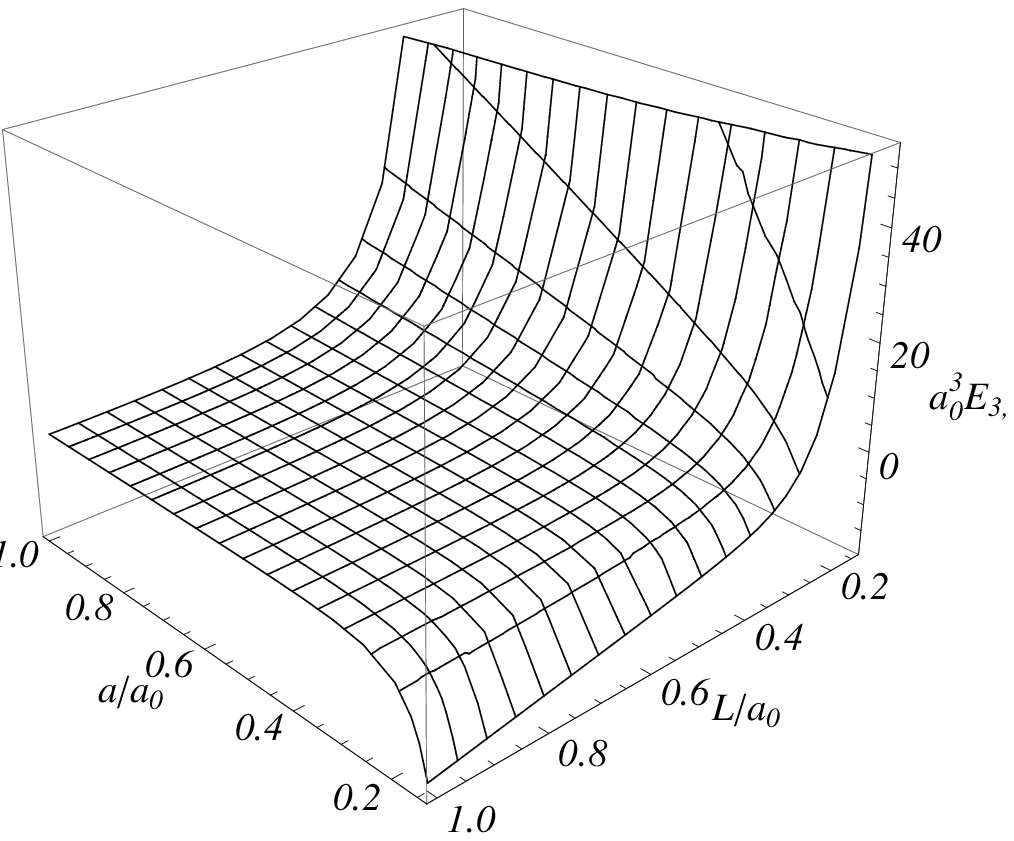,width=7cm,height=6.cm} & \quad %
\epsfig{figure=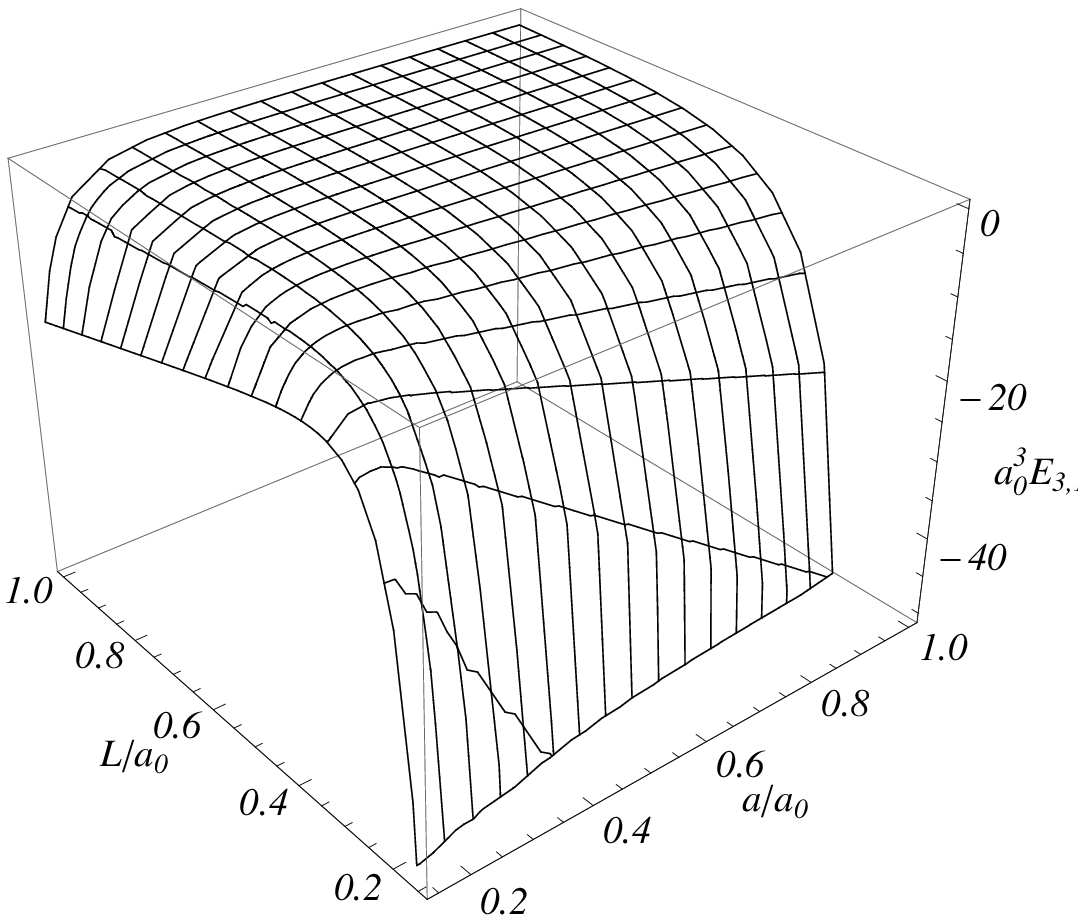,width=7cm,height=6cm}%
\end{tabular}%
\end{center}
\caption{The Casimir energy of a massless fermionic field in 4-dimensional
space with topology $R^{3}\times S^{1}$ as a function of the inter-plate
distance and the length of the compact dimension. The left/right panel
corresponds to untwisted/twisted fields. }
\label{fig1}
\end{figure}

For the massless case these formulae are simplified to%
\begin{eqnarray}
E_{D-1,1}^{(0)} &=&\frac{NL^{-D}}{\pi ^{(D+1)/2}}\Gamma
((D+1)/2)\sum_{n=1}^{\infty }\frac{\cos (2\pi n\alpha )}{n^{D+1}},  \notag \\
\Delta E_{D-1,1} &=&\frac{(2\pi )^{-D/2}N}{(2a)^{D-1}}\sum_{l=-\infty
}^{+\infty }\sum_{n=1}^{\infty }\frac{(-1)^{n}}{n^{D}}f_{D/2}(4\pi
n|l+\alpha |a/L),  \label{Espm0}
\end{eqnarray}%
and the single plate part vanishes. In figure \ref{fig1} we have presented
the Casimir energy $E_{D-1,1}$ for a massless fermionic field in the
simplest Kaluza-Klein model with $D=4$ as a function of the inter-plate
distance and the length of the internal space measured in units of a fixed
length $a_{0}$. The left panel corresponds to the untwisted field ($\alpha
=0 $) and the right one is for the twisted field ($\alpha =1/2$). For large
inter-plate separations the pure topological part dominates and the Casimir
energy is a linear function of $a$. At small distances the interaction part
is dominant and the Casimir energy behaves as $a^{-D}$.

\begin{figure}[tbph]
\begin{center}
\epsfig{figure=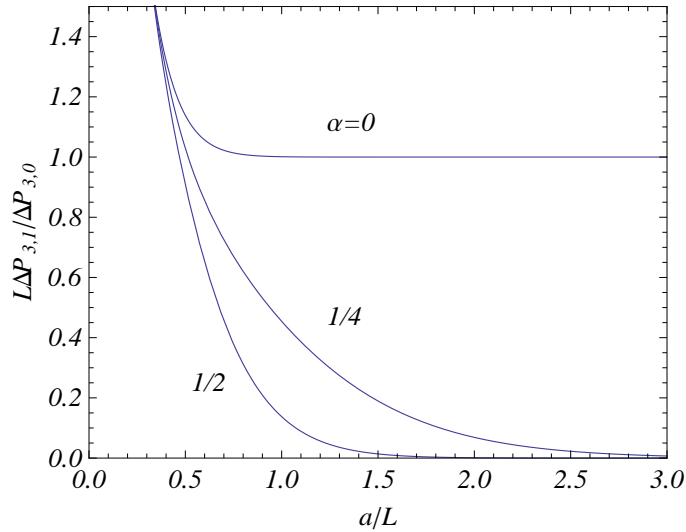,width=9.cm,height=7.cm}
\end{center}
\caption{The ratio of the fermionic Casimir force for two parallel plates in
the space with topology $R^{3}\times S^{1}$ to the standard Casimir force in
$R^{3}$, for a massless field, as a function of $a/L$. The values on each of
the curves correspond to those of the parameter $\protect\alpha $.}
\label{fig2}
\end{figure}

In the special case under consideration for the interaction part of the
Casimir force we have the formula
\begin{equation}
\Delta P_{D-1,1}=-\frac{2(4\pi )^{-(D-1)/2}N}{\Gamma ((D-1)/2)L_{D}}%
\sum_{l=-\infty }^{+\infty }\int_{m_{l}}^{\infty }dz\frac{%
z^{2}(z^{2}-m_{l}^{2})^{(D-3)/2}}{\frac{z+m}{z-m}e^{-2az}+1}.
\label{DeltPsp}
\end{equation}%
In the massless case this formula takes the form
\begin{equation}
\Delta P_{D-1,1}=-\frac{2N}{(2\pi )^{D/2}L}\sum_{l=-\infty }^{+\infty
}\sum_{n=1}^{\infty }(-1)^{n}\frac{f_{D/2}(y)-f_{D/2+1}(y)}{(2an)^{D}}%
|_{y=4\pi n|l+\alpha |a/L}.  \label{DPspm0}
\end{equation}%
In figure \ref{fig2} we have plotted the ratio $L\Delta P_{3,1}/\Delta
P_{3,0}$ versus $a/L$ for different values of the parameter $\alpha $. As it
already has been explained before, only in the case of untwisted field the
Casimir force at large separations tends to the corresponding force (up to
the factor related to the number of polarizations) for the model where the
compactified dimensions are absent. For other cases the force is
exponentially suppressed at large separations which is clearly seen in
figure \ref{fig2}.

\section{Applications to finite-length nanotubes}

\label{sec:Nanotubes}

For a number of planar condensed matter systems the fermionic excitations in
the long-wavelength regime are described by the Dirac-like model. A well
known example is the graphene. In this section we specify the general
results given above for the electrons on a graphene sheet rolled into a
cylindrical shape (carbon nanotube). The carbon nanotube is characterized by
its chiral vector $\mathbf{C}_{h}=n_{w}\mathbf{a}_{1}+m_{w}\mathbf{a}_{2}$,
where $\mathbf{a}_{1}$ and $\mathbf{a}_{2}$ are the basis vectors of the
hexagonal lattice of graphene and $n_{w}$, $m_{w}$ are integers. The
circumference length of the nanotube is given by $L=|\mathbf{C}_{h}|=a_{%
\mathrm{g}}\sqrt{n_{w}^{2}+m_{w}^{2}+n_{w}m_{w}}$, with $a_{\mathrm{g}}=|%
\mathbf{a}_{1}|=|\mathbf{a}_{2}|=2.46\mathring{A}$ being the lattice
constant. Zigzag nanotubes correspond to the special case $\mathbf{C}%
_{h}=(n_{w},0)$, and for armchair nanotubes one has $\mathbf{C}%
_{h}=(n_{w},n_{w})$. All other cases correspond to chiral nanotubes. The
electron properties of carbon nanotubes can be either metallic or
semiconductor-like depending on the manner the cylinder is obtained from the
graphene sheet. In the case $n_{w}-m_{w}=3q_{w}$, $q_{w}\in Z$, the nanotube
will be metallic and in the case $n_{w}-m_{w}\neq 3q_{w}$ the nanotube will
be semiconductor with an energy gap inversely proportional to the diameter.
In particular, the armchair nanotube is metallic and the $(n_{w},0)$ zigzag
nanotube is metallic if and only if $n_{w}$ is an integer multiple of 3.

The electronic band structure of a carbon nanotube close to the Dirac points
shows a conical dispersion $E(\mathbf{k})=v_{\mathrm{F}}|\mathbf{k}|$, where $\mathbf{%
k}$ is the momentum measured relatively to the Dirac points and $v_{\mathrm{F%
}}$ represents the Fermi velocity which plays the role of speed of light.
The corresponding low-energy excitations can be described by a pair of
two-component spinors, which are composed of the Bloch states residing on
the two different sublattices of the honeycomb lattice of the graphene
sheet. The corresponding Fermi velocity is given by $v_{\mathrm{F}}=3ta/2$ ($%
v_{\mathrm{F}}\approx 10^{8}\mathrm{cm/s}$ in graphene), where $t$ is the
nearest neighbor hopping energy. The Dirac-like model is valid provided that
the cylinder circumference is much larger than the interatomic spacing. For
typical nanotubes the corresponding ratio can be between 10 and 20 and this
approximation is adequate \cite{Sait98,Seme84}. In the case under
consideration $D=2$ and we have the spatial topology $R^{1}\times S^{1}$
with the compactified dimension of the length $L$. We will assume that the
nanotube has finite length $a$. As the $D=2$ Dirac field lives on the
cylinder surface it is natural to impose bag boundary conditions (\ref%
{BagCond}) on the cylinder edges which insure the zero fermion flux through
these edges. The additional confinement of the electrons along the tube axis
leads to the change of the ground state energy. The corresponding
expressions for the Casimir energy and force are obtained from the formulae
of the previous section taking $D=2$. Here, by taking into account that in
the presence of an external magnetic field an effective mass term is
generated for the fermionic excitations, we consider the general case of
massive spinor field. The formulae for a massless case, appropriate for
carbon nanotubes in the absence of external fields, will be given separately.

In order to specify the boundary condition on the fermionic field along the
compactified dimension, we note that for the $(n_{w},m_{w})$ nanotube the
phase factor in the wavefunction has the form $e^{i[m_{1}+(n_{w}-m_{w})/3]%
\varphi }$, where $\varphi $ is the angular coordinate along the compact
dimension and $m_{1}$ is an integer. From here it follows that for metallic
nanotubes the periodic boundary condition ($\alpha =0$) is realized. For
semiconductor nanotubes, depending on the chiral vector, there are two
classes of inequivalent boundary conditions corresponding to $\alpha =1/3$ ($%
n_{w}-m_{w}=3q_{w}+2$) and $\alpha =2/3$ ($n_{w}-m_{w}=3q_{w}+1$). In the
expressions for the pure topological parts of the Casimir energy and force
the phase $\alpha $ appears in the form $\cos (2\pi n\alpha )$ and, hence,
these quantities are the same for $\alpha =1/3$ and $\alpha =2/3$. As the
boundary induced parts have the structure $\sum_{l=-\infty }^{+\infty
}f(|l+\alpha |)$, the same property holds for these parts.

In the case $D=2$, the general formulae for the separate parts of the
Casimir energy from the previous section take the form ($N=2$)%
\begin{eqnarray}
E_{1,1}^{(0)} &=&\frac{1}{\pi L^{2}}\sum_{n=1}^{\infty }(1+mnL)\cos (2\pi
n\alpha )\frac{e^{-mnL}}{n^{3}},  \notag \\
E_{1,1}^{(1,c)} &=&\frac{m^{2}L}{2\pi }\sum_{n=1}^{\infty }\cos (2\pi
n\alpha )\sum_{l=1}^{\infty }\frac{2^{-l/2}(-1)^{l}}{\Gamma (l/2+1)}%
f_{l/2-1}(nLm),  \label{ECN} \\
\Delta E_{1,1} &=&-\frac{1}{\pi }\sum_{l=-\infty }^{+\infty
}\int_{0}^{\infty }dz\ln \left( 1+\frac{\sqrt{z^{2}+m_{l}^{2}}-m}{\sqrt{%
z^{2}+m_{l}^{2}}+m}e^{-2a\sqrt{z^{2}+m_{l}^{2}}}\right) .  \notag
\end{eqnarray}%
For the Casimir force acting on the edges of the tube we have%
\begin{eqnarray}
\;P_{1,1} &=&-\frac{1}{\pi L^{3}}\sum_{n=1}^{\infty }(1+mnL)\cos (2\pi
n\alpha )\frac{e^{-mnL}}{n^{3}}  \notag \\
&&-\frac{2}{\pi L}\sum_{l=-\infty }^{+\infty }\int_{0}^{\infty }dz\,z\left(
\frac{\sqrt{z^{2}+m_{l}^{2}}+m}{\sqrt{z^{2}+m_{l}^{2}}-m}e^{2a\sqrt{%
z^{2}+m_{l}^{2}}}+1\right) ^{-1}.  \label{PCN}
\end{eqnarray}

In the massless case for the total Casimir energy and the stresses we find
the formulae%
\begin{eqnarray}
E_{1,1} &=&\frac{a}{\pi L^{2}}\sum_{n=1}^{\infty }\frac{\cos (2\pi n\alpha )%
}{n^{3}}-\frac{1}{2\pi a}\sum_{l=-\infty }^{+\infty }\sum_{n=1}^{\infty }%
\frac{(-1)^{n}}{n^{2}}f_{1}(4\pi n|l+\alpha |a/L),  \notag \\
P_{1,1} &=&-\frac{1}{\pi L^{3}}\sum_{n=1}^{\infty }\frac{\cos (2\pi n\alpha )%
}{n^{3}}-\frac{1}{2\pi a^{2}L}\sum_{l=-\infty }^{+\infty }\sum_{n=1}^{\infty
}(-1)^{n}\frac{f_{1}(y)-f_{2}(y)}{n^{2}}|_{y=4\pi n|l+\alpha |a/L}.
\label{EPm0CN}
\end{eqnarray}%
The corresponding expressions for the Casimir energy and force in finite
length cylindrical nanotubes are obtained from (\ref{EPm0CN}) with
additional factor 2 which takes into account the presence of two
sublattices. In standard units the factor $\hbar v_{\mathrm{F}}$ appears as
well. So, for the Casimir force acting per unit length of the edge of a
carbon nanotube one has: $P^{\mathrm{(CN)}}=2\hbar v_{\mathrm{F}}P_{1,1}$,
where $P_{1,1}$ is given by Eq. (\ref{EPm0CN}). For long tubes, $a/L\gg 1$,
the first term on the right is dominant and we have $P^{\mathrm{(CN)}%
}\approx -0.765\hbar v_{\mathrm{F}}/L^{3}$ for metallic nanotubes and $P^{%
\mathrm{(CN)}}\approx 0.34\hbar v_{\mathrm{F}}/L^{3}$ for semiconducting
ones. In the limit $a/L\ll 1$ the interaction part is dominant. In the
leading order the Casimir force do not depend on the chirality and one has $%
P^{\mathrm{(CN)}}\approx -0.144\hbar v_{\mathrm{F}}/a^{3}$. In figure \ref%
{fig3} we have plotted the Casimir forces acting on the edges of metallic
(left panel) and semiconducting-type (right panel) carbon nanotube as
functions of the tube length for different values of the fermion mass. As it
is seen, for metallic nanotubes these forces are always attractive, whereas
for semiconducting-type ones they are attractive for small lengths and
repulsive for large lengths.
\begin{figure}[tbph]
\begin{center}
\begin{tabular}{cc}
\epsfig{figure=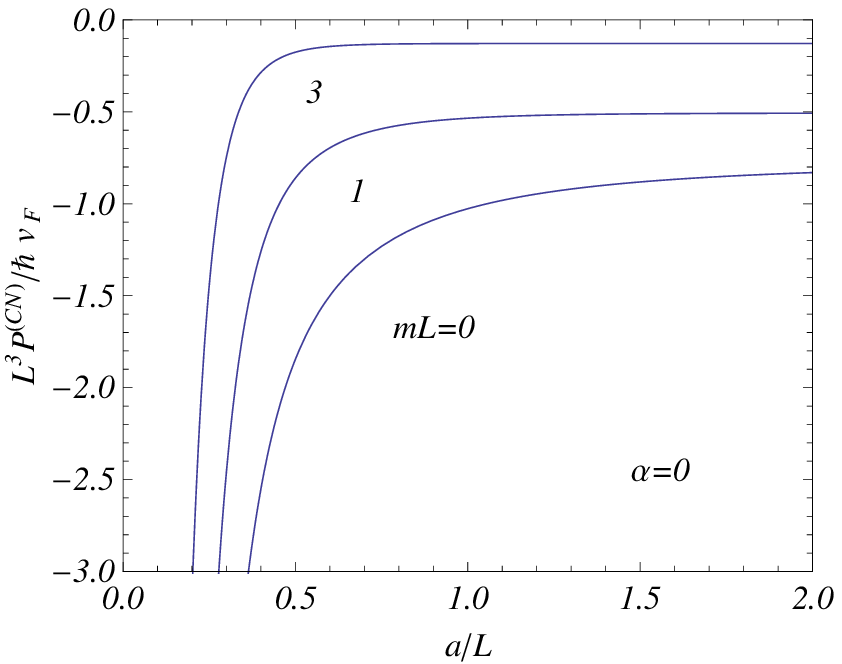,width=7cm,height=6.cm} & \quad %
\epsfig{figure=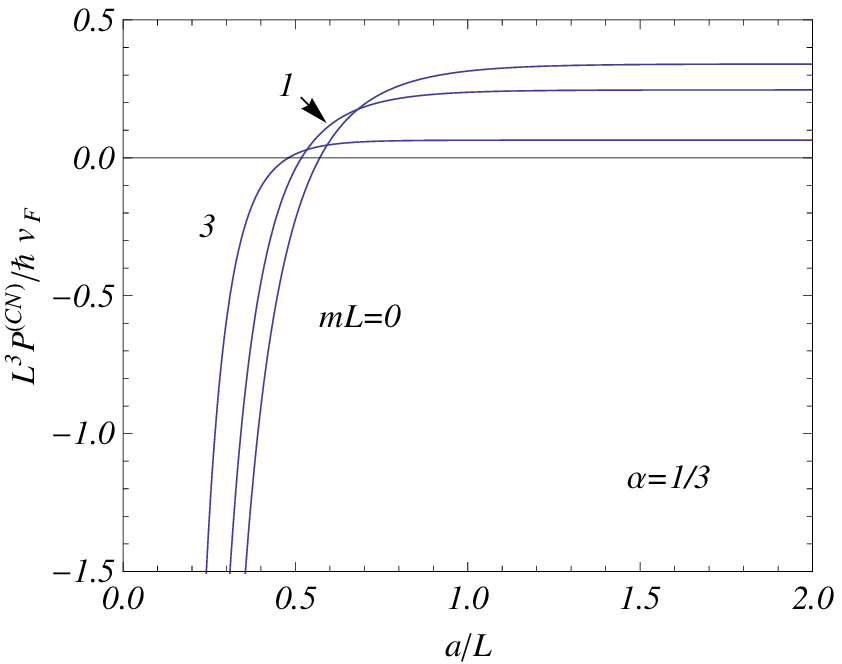,width=7cm,height=6cm}%
\end{tabular}%
\end{center}
\caption{The fermionic Casimir forces acting on the edges of the metallic
(left panel) and semiconducting-type (right panel) nanotubes as functions of
the tube length for different values of the field mass.}
\label{fig3}
\end{figure}

In the discussion above we have considered bag boundary conditions
on the edges of the nanotube. The periodicity conditions along the
axis correspond to the toroidal compactification of the carbon
nanotubes. The Casimir energies in toroidal nanotubes are
investigated in Ref. \cite{Bell09}, where it was shown that the
toroidal compactification of a cylindrical nanotube along its axis
increases the Casimir energy for periodic boundary conditions and
decreases the Casimir energy for the semiconducting-type
compactifications. Recently, in the last paper of Ref.
\cite{CasNano}, the Casimir interaction between two plates
resulting from the quantum fluctuations of the bulk
electromagnetic field is investigated with one plate being
graphene described the Dirac model and the other one being ideal
conductor. The interaction of the electromagnetic field with the
fermion field confined on the graphene sheet is equivalent to
imposing boundary condition for the electromagnetic field. At
large separations the corresponding force is proportional to the
fine structure constant and falls off as the inverse cube of
distance between the plates.

\section{Conclusion}

\label{sec:Conclusion}

We have investigated the effect of compact spatial dimensions on the Casimir
energy and force for a massive fermionic field in the geometry of two
parallel plates on which the field obeys MIT bag boundary condition. Along
the compact dimensions we have assumed periodicity conditions (\ref{BC})
with constant phases $\alpha _{l}$. The eigenvalues of the wave-vector
component normal to the plates are roots of transcendental equation (\ref%
{kpvalues}). By applying the Abel-Plana-type summation formula to the
corresponding series in the mode-sum for the vacuum energy in the region
between the plates, we have explicitly extracted, in a cut-off independent
way, the pure topological part and the contributions induced by the single
plates. The surface divergences in the Casimir energy are contained in the
single plate components only and the remaining interaction part is finite
for all nonzero inter-plate distances. The latter is given by Eq.~(\ref%
{DeltaE2}) for a massive field and by Eq. (\ref{DeltaEm0}) in the massless
case. The interaction part of the Casimir energy is always negative. We have
decomposed the single plate part in the vacuum energy into two terms: the
first one is the Casimir energy for a single plate in the trivial topology $%
R^{D}$ and the second one is the topological part. The second term is
cutoff-independent and in this way the renormalization procedure is reduced
to that for the plate in topology $R^{D}$.

The Casimir forces between the plates have been considered in section~\ref%
{sec:CasForce}. Single plate parts in the Casimir energy do not depend on
the plates separation and do not contribute to the force. \ For the region
between the plates the forces are presented as the sum of topological and
interaction parts. In the situations where the quantum field lives on both
sides of the plate, the topological parts are the same on the left and right
sides and the effective force is determined by the interaction part only.
The latter is given by formulae (\ref{DeltP}) and (\ref{DeltaPm0}) for the
massive and massless fields respectively. With independence of the lengths
of compact dimensions and the phases in the periodicity conditions, the
corresponding force is attractive and is a monotonic function of the
distance. When the field is confined in the region between the plates only
the topological part contributes to the resulting force and it dominates at
large separations between the plates. In dependence of the phases in the
periodicity conditions, the corresponding forces can be either attractive or
repulsive. In particular, for untwisted fields the Casimir forces are
attractive for all separations and for twisted fields these forces are
attractive for small distances and repulsive at large distances. For small
separations the interaction part dominates and the Casimir force is
attractive. For small values of the size of the compact subspace and in
models where the zero mode along the internal space is present, the main
contribution to the Casimir force comes from this mode and the contributions
of the nonzero modes are exponentially suppressed. In this limit, to leading
order we recover the standard result for the Casimir force between two
plates in $(p+2)$-dimensional Minkowski spacetime. When the zero mode is
absent, the Casimir forces are exponentially suppressed in the limit of
small size of the internal space.

In section \ref{sec:Zeta} we have evaluated the Casimir energy by using an
alternative method based on the generalized zeta function technique. With
the combination of the extended Chowla--Selberg formula, this allowed us to
present the topological part for the geometry of a single plate in an
alternative form given by formula (\ref{E1c}). As an illustration of the
general results, in Sect.~\ref{sec:SpCase} we have considered a special
model with a single compact dimensions. In section~\ref{sec:Nanotubes} we
specify the general formulae for the model with $D=2$. This model may be
used for the evaluation of the Casimir energy and force within the framework
of the Dirac-like theory for the description of the electronic states in
carbon nanotubes where the role of speed of light is played by the Fermi
velocity. The pure topological part of the Casimir energy is positive for
metallic cylindrical nanotubes and is negative for semiconducting ones. For
finite-length carbon nanotubes the Casimir forces acting on the tube edges
are always attractive for metallic nanotubes, whereas for
semiconducting-type ones they are attractive for small lengths and repulsive
for large lengths.

\section*{Acknowledgments}

A.A.S. was supported by the Armenian Ministry of Education and Science Grant
No. 119.

\end{document}